\documentclass[aps,prb,twocolumn,showpacs,superscriptaddress,longbibliography]{revtex4-2}
\usepackage{amsfonts}
\usepackage{amsmath}
\usepackage{graphicx}
\usepackage{bm}
\usepackage{amssymb}
\usepackage{dcolumn}
\usepackage{color}
\usepackage{multirow}
\usepackage{booktabs}
\usepackage{diagbox}
\usepackage[colorlinks,
    linkcolor=blue,
    anchorcolor=blue,
    citecolor=blue,
    urlcolor=blue,
]{hyperref}
\usepackage{verbatim}

\begin{document}

    \title{High temperature ferrimagnetic semiconductors by spin-dependent doping in high temperature antiferromagnets}

    \author{Jia-Wen Li}
    \affiliation{Kavli Institute for Theoretical Sciences, University of Chinese Academy of Sciences, Beijng 100049, China}

    \author{Gang Su}
    \email{gsu@ucas.ac.cn}
    \affiliation{Kavli Institute for Theoretical Sciences, University of Chinese Academy of Sciences, Beijng 100049, China}
    \affiliation{CAS Center for Excellence in Topological Quantum Computation, University of Chinese Academy of Sciences, Beijng 100190, China}
    \affiliation{Physical Science Laboratory, Huairou National Comprehensive Science Center, Beijing 101400, China}
    \affiliation{School of Physical Sciences, University of Chinese Academy of Sciences, Beijng 100049, China}

    \author{Bo Gu}
    \email{gubo@ucas.ac.cn}
    \affiliation{Kavli Institute for Theoretical Sciences, University of Chinese Academy of Sciences, Beijng 100049, China}
    \affiliation{CAS Center for Excellence in Topological Quantum Computation, University of Chinese Academy of Sciences, Beijng 100190, China}
    \affiliation{Physical Science Laboratory, Huairou National Comprehensive Science Center, Beijing 101400, China}

    \begin{abstract}
        To realize room temperature ferromagnetic (FM) semiconductors is still a challenge in spintronics.
        Many antiferromagnetic (AFM) insulators and semiconductors with high Neel temperature $T_N$ are obtained in experiments, such as LaFeO$_3$, BiFeO$_3$, etc.
        High concentrations of magnetic impurities can be doped into these AFM materials, but AFM state with very tiny net magnetic moments was obtained in experiments, because the magnetic impurities were equally doped into the spin up and down sublattices of the AFM materials.
        Here, we propose that the effective magnetic field provided by a FM substrate could guarantee the spin-dependent doping in AFM materials, where the doped magnetic impurities prefer one sublattice of spins, and the ferrimagnetic (FIM) materials are obtained.
        To demonstrate this proposal, we study the Mn-doped AFM insulator LaFeO$_3$ with FM substrate of Fe metal by the density functional theory (DFT) calculations.
        It is shown that the doped magnetic Mn impurities prefer to occupy one sublattice of AFM insulator, and introduce large magnetic moments in La(Fe,Mn)O$_3$.
        For the AFM insulator LaFeO$_3$ with high $T_N$ = 740 K, several FIM semiconductors with high Curie temperature $T_C >$ 300 K and the band gap less than 2 eV are obtained by DFT calculations, when 1/8 or 1/4 Fe atoms in LaFeO$_3$ are replaced by the other 3d, 4d transition metal elements.
        The large magneto-optical Kerr effect (MOKE) is obtained in these LaFeO$_3$-based FIM semiconductors.
        In addition, the FIM semiconductors with high $T_C$ are also obtained by spin-dependent doping in some other AFM materials with high $T_N$, including BiFeO$_3$, SrTcO$_3$, CaTcO$_3$, etc.
        Our theoretical results propose a way to obtain high $T_C$ FIM semiconductors by spin-dependent doping in high $T_N$ AFM insulators and semiconductors.
    \end{abstract}
    \maketitle


    \section{Introduction}
    In spintronics, it is still a challenge in experiments to realize room temperature ferromagnetic (FM) semiconductors.
    The Curie temperature $T_C$ of intrinsic two- and three-dimensional FM semiconductors are still far below the room temperature \cite{Huang2017,Lee2017,Cai2019,Chu2019,Zhang2019,Achinuq2021,Lee2021,Baltzer1966,Gong2017}, which largely limit their applications.

    Doping is an effective approach to control the physical properties of materials.
    By doping a small amount of magnetic impurities into non-magnetic semiconductors, the magnetic properties of the materials can be dramatically improved, these materials are called dilute magnetic semiconductors (DMS) \cite{Ohno1998,Jungwirth2006,Sato2010,Dietl2014,Zhao2022,Dong2022,Huang2021,Kalita2023}.
    For the classic DMS (Ga,Mn)As, its highest $T_C$ can reach 200 K \cite{Chen2011}.
    High $T_C$ DMSs have been reported in recent experiments, such as  $T_C$ = 230 K in (Ba,K)(Zn,Mn)$_2$As$_2$ with 15\% doping of Mn \cite{Zhao2013,Zhao2014}, $T_C$ = 340 K in (Ga, Fe)Sb with 25\% doping of Fe \cite{Tu2016}, $T_C$ = 385 K in (In, Fe)Sb with 35\% doping of Fe \cite{Tu2019}, T$_C$ = 280 K in (Si$_{0.25}$Ge$_{0.75}$, Mn) with 5\% doping of Mn \cite{Wang2020}, etc.

    In contrast to DMS, there are also some studies on the magnetic impurities doped antiferromagnetic (AFM) insulators and semiconductors in experiments.
    Some AFM insulators and semiconductors with high Neel temperature $T_N$ have been obtained experimentally, as shown in Table \ref{table_AFM} \cite{Marezio1971,Yakel1955,Koehler1957,Manzoor2018,Silva2011,Rodriguez2011,Avdeev2011,Sawatzky1984,Marynowski1999,Yanagi2009,Beleanu2013,Emery2010,Kriegner2016,Wijnheijmer2012,Beleanu2013,Brockhouse1953,Abdullah2014}.
    Being a high $T_N$ AFM insulator, LaFeO$_3$ has attracted a lot of attentions due to its interesting properties.
    LaFeO$_3$ has a perovskite structure with chemical formula of ABO$_3$ \cite{Marezio1971,Yakel1955,Koehler1957}.
    A high $T_N$ = 740 K has been observed in LaFeO$_3$ \cite{Koehler1957}, where the magnetic ground state is G-AFM with intralayer and interlayer AFM order.
    LaFeO$_3$ has a large optical band gap of 2.05 to 2.51 eV in experiments \cite{Sasikala2017,Manzoor2018}.
    Room temperature ferroelectricity of LaFeO$_3$ has also been observed \cite{Acharya2010}.
    In addition, the doped LaFeO$_3$ has also been studied, such as (La, X)FeO$_3$ with X = Sr \cite{Takano1981}, Al \cite{Acharya2011}, Bi \cite{Ahmed2015,Yao2020}, Ca \cite{Bidrawn2008}, Ba \cite{Bidrawn2008}, and La(Fe, D)O$_3$ with D = Mo \cite{Jana2019}, Ni \cite{Idrees2011}, Cr \cite{Azad2005,Rodrigues2020,Xia2022,Paul_Blessington_Selvadurai_2015}, Ti \cite{Phokha2015,Sasikala2018,Sasikala2017}, Zn \cite{Bhat2013,Manzoor2018}, Cu \cite{Dogdibegovic2016}, Mn \cite{Bhame2005}, Mg \cite{Triyono2020}, Co \cite{Gu2021} etc.
    It shows a high tolerance to impurities, the doping concentration at both La and Fe sites could reach to about 50\%.
    Some magnetic impurities doped AFM insulators and semiconductors with high $T_N$ are shown in Table \ref{dop_exp}.
    The experimental studies of La(Fe$_{1-x}$D$_x$)O$_3$  \cite{Jana2019,Idrees2011,Azad2005,Rodrigues2020,Xia2022,Paul_Blessington_Selvadurai_2015,Phokha2015,Sasikala2018,Sasikala2017,Bhat2013,Manzoor2018,Dogdibegovic2016,Bhame2005}, Bi(Fe$_{1-x}$D$_x$)O$_3$ \cite{Sosnowska2002,Khajonrit2018,Sui2015,Jun2005,Kharel2008,Mukherjee2012} and (Ni$_{1-x}$D$_x$)O \cite{Panigrahi2020,Manna2008,Layek2016,Rahman2018} have shown very tiny net magnetic moments, although the high concentrations of magnetic impurities can be realized.

\begin{figure}[phbpt]
        \centering
        \includegraphics[scale=0.38,angle=0]{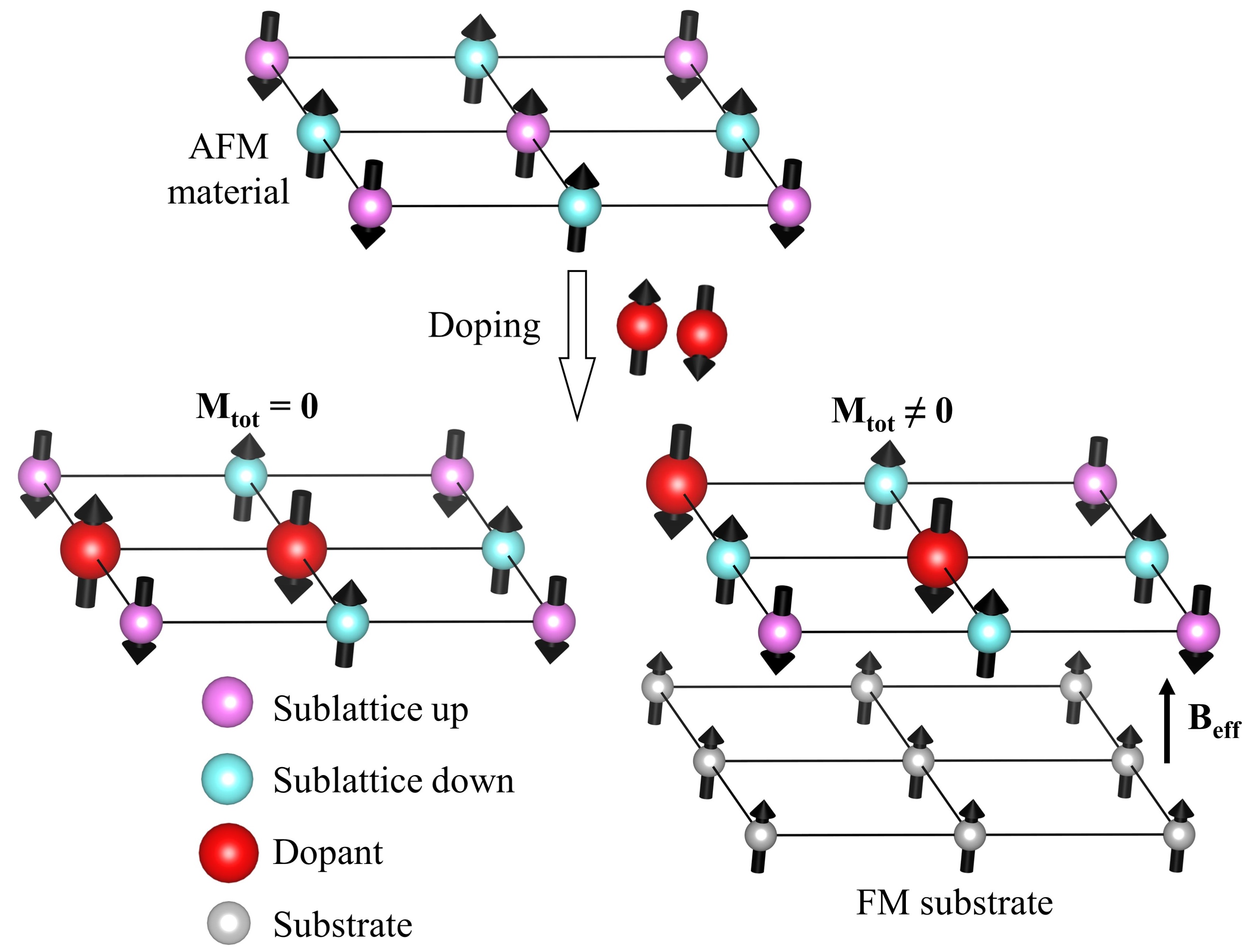}\\
        \caption{Schematic diagram of spin-independent doping (left) with zero net magnetic moment and spin-dependent doping (right) with non-zero net magnetic moment, for the antiferromagnetic (AFM) materials doped with magnetic impurities. }\label{fig1}
    \end{figure}

    As shown in Table \ref{dop_exp}, there is an increase of net magnetic moment in AFM materials after doping, which was explained as the formation of clusters \cite{Azad2005,Paul_Blessington_Selvadurai_2015,Rodrigues2020,Bhat2013}, enhancement of interface effects \cite{Paul_Blessington_Selvadurai_2015,Phokha2015,Sasikala2017}, change of magnetic coupling \cite{Xia2022,Rodrigues2020,Paul_Blessington_Selvadurai_2015,Dogdibegovic2016}, etc.
    However, their net magnetic moment is still negligible, which can be understood from the symmetry of spin up and down sublattices of AFM host materials.
    As shown in Fig. \ref{fig1}, magnetic impurities were equally doped into the spin up and down sublattices of the AFM materials, resulting in zero net magnetic moment.
    On the other hand, as shown in Table \ref{dop_exp}, only a few theoretical studies focus on the magnetic impurities doped AFM insulators and semiconductors, and nearly have not discussed the theoretical results of magnetic properties, such as $T_N$
    \cite{Tariq2022,Zhou2021,Azouzi2021,Lu2017,Rong2016,Egbo2020}.
    Is there a way to break the symmetry of spin up and down sublattices of AFM host materials?

    In this paper, we propose that the effective magnetic field from the FM substrate can break the symmetry of spin up and down sublattices and make the spin-dependent doping possible in AFM materials, as schematically shown in Fig. \ref{fig1}.
    To demonstrate our proposal, we study the Mn-doped AFM insulator LaFeO$_3$ with FM substrate of Fe metal by the density functional theory (DFT) calculations.
    The calculation results for the supercell La(Fe,Mn)O$_3$/bcc-Fe show that the doped magnetic Mn impurities prefer to occupy one sublattice of AFM insulator, and introduce large magnetic moments in La(Fe,Mn)O$_3$.
    By this way, some ferrimagnetic (FIM) semiconductors with Curie temperature $T_C$ above room temperature are predicted for La(Fe$_{1-x}$D$_x$)O$_3$ with D = 3d, 4d transition metal impurities and x = 0.125 and 0.25.
    In addition, La(Fe$_{0.75}$D$_{0.25}$)O$_3$ shows large magneto-optical Kerr effect.
    The variation of $T_C$ in the FIM La(Fe$_{1-x}$D$_x$)O$_3$ as a function of elements D can be well understood by a formula of mean-field theory.
    Our results propose a way to obtain high temperature FIM semiconductors by spin-dependent doping in high temperature AFM insulators and semiconductors.

    \begin{table}[btht]
        \setlength{\tabcolsep}{2mm}
        \caption{{Some antiferromagnetic (AFM) insulators and semiconductors with high Neel temperatrue $T_N$ in experiments.}}
        {
            \scalebox{1}

            {
                \begin{tabular}{cccc}
                    \hline
                    \hline
                    \makebox[0.09\textwidth][c]{AFM materials}
                    &
                    \makebox[0.09\textwidth][c]{$T_N$ (K)}
                    &
                    \makebox[0.09\textwidth][c]{Gap (eV)}
                    &
                    \makebox[0.09\textwidth][c]{Experiments}
                    \\
                    \hline
                    LaFeO$_3$ & 740 & 2.5 &
                    Ref. \cite{Yakel1955,Marezio1971,Koehler1957,Manzoor2018} \\
                    BiFeO$_3$ & 640 & 2.5 &
                    Ref. \cite{Silva2011} \\
                    SrTcO$_3$ & 1023 & 1.5 &
                    Ref. \cite{Rodriguez2011} \\
                    CaTcO$_3$ & 850 & 2.2 &
                    Ref. \cite{Avdeev2011} \\
                    NiO & 525 & 3.2 &
                    Ref. \cite{Sawatzky1984,Marynowski1999} \\
                    LaOMnP & 375 & 1.4 &
                    Ref. \cite{Yanagi2009} \\
                    LaOMnAs & 317 & 0.4 &
                    Ref. \cite{Beleanu2013,Emery2010} \\
                    MnTe & 307 & 1.4 &
                    Ref. \cite{Kriegner2016} \\
                    LiMnAs & 374 & 0.2 &
                    Ref. \cite{Wijnheijmer2012,Beleanu2013} \\
                    Cr$_2$O$_3$ & 340 & 3.3 &
                    Ref. \cite{Brockhouse1953,Abdullah2014} \\
                    \hline
                    \hline

                \end{tabular}
            }}

        \label{table_AFM}
    \end{table}

    \section{Method}
    Our calculations were based on the DFT as implemented in the Vienna ab initio simulation package (VASP) \cite{Kresse1996}.
    The exchange-correlation potential is described by the Perdew-Burke-Ernzerhof (PBE) form with the generalized gradient approximation (GGA) \cite{Perdew1996}.
    The electron-ion potential is described by the projector-augmented wave (PAW) method \cite{Bloechl1994}.
    We carried out the calculation of GGA + U with U = 4 or 2 eV for 3d or 4d elements, respectively.
    The plane-wave cutoff energy is set to be 500 eV.
    The 4$\times$4$\times$1, 4$\times$4$\times$3 and 2$\times$4$\times$3 $\Gamma$ center k-point meshed were used for the Brillouin zone (BZ) sampling for supercells of La(Fe$_{0.75}$D$_{0.25}$)O$_3$/bcc-Fe, La(Fe$_{0.75}$D$_{0.25}$)O$_3$ and La(Fe$_{0.875}$D$_{0.125}$)O$_3$, respectively.
    The structures of all materials were fully relaxed, where the convergence precision of energy and force were $10^{-6}$ eV and $10^{-2}$ eV/Å, respectively.
    The van der Waals effect is include with DFT-D3 method \cite{Grimme2010}.
    The Wannier90 code was used to construct a tight-binding Hamiltonian to calculate the Kerr rotation angle \cite{Mostofi2008,Mostofi2014}.
    The Heisenberg-type Monte Carlo simulation was performed on 10$\times$10$\times$10 and 8$\times$8$\times$8 lattice with 4000 and 4096 magnetic points for La(Fe$_{0.75}$D$_{0.25}$)O$_3$ and La(Fe$_{0.875}$D$_{0.125}$)O$_3$, respectively.
    More than 8$\times$10$^4$ steps were carried for each temperature, and the last one-thirds steps were used to calculate the temperature-dependent physical quantities.

    \begin{table*}[thb]
        \setlength{\tabcolsep}{2mm}
        \caption{
            Some magnetic impurities doped AFM insulators and semiconductors with high $T_N$.
            x is the doping concentration. $<$M$>$ is the average magnetic moment per magnetic atom in unit of $\mu_B$.
            RT means $T_N$ is above room temperature, and y and n denotes yes and no, respectively.
            In addition, / denote that the related property is not discussed in the references.
        }
        {
            \scalebox{1}

            {
                \begin{tabular}{|c|c|c|c|c|c|c|c|c|c|}
                    \hline
                    \multirow{2}{*}{\makebox[0.18\textwidth][c]{\diagbox{Materials}{Properties}}}
                    &
                    \multirow{2}{*}{\makebox[0.06\textwidth][c]{D}} 
                    &
                    \multicolumn{5}{c|}{\makebox[0.30\textwidth][c]{Experiments}} 
                    &
                    \multicolumn{3}{c|}{\makebox[0.18\textwidth][c]{Theories}}
                    \\
                    \cline{3-10}
                    & & \makebox[0.06\textwidth][c]{x}
                    & \makebox[0.06\textwidth][c]{$<$M$>$}
                    & \makebox[0.06\textwidth][c]{$T_N$}
                    & \makebox[0.06\textwidth][c]{Gap}
                    & \makebox[0.06\textwidth][c]{Ref}
                    & \makebox[0.06\textwidth][c]{x}
                    & \makebox[0.06\textwidth][c]{Gap}
                    & \makebox[0.06\textwidth][c]{Ref}
                    \\
                    \hline
                    \multirow{10}{*}{La(Fe$_{1-x}$D$_x$)O$_3$}
                    & Mo & 0.25 & ~1$\times$10$^{-2}$
                    & \multirow{6}{*}{RT}
                    & \multirow{8}{*}{y}
                    & \cite{Jana2019}
                    & \multicolumn{3}{c|}{}
                    \\
                    & Zn & 0.30 & ~1$\times$10$^{-4}$
                    & &
                    & \cite{Bhat2013,Manzoor2018}
                    & \multicolumn{3}{c|}{}
                    \\
                    & Ti & 0.20 & ~2$\times$10$^{-3}$
                    & &
                    & \cite{Phokha2015,Sasikala2018,Sasikala2017}
                    & \multicolumn{3}{c|}{}
                    \\
                    & Ni & 0.30 & ~1$\times$10$^{-2}$
                    & &
                    & \cite{Idrees2011}
                    & \multicolumn{3}{c|}{}
                    \\
                    & Cu & 0.20 & /
                    & &
                    & \cite{Dogdibegovic2016}
                    & \multicolumn{3}{c|}{}
                    \\
                    \cline{8-10}
                    & Cr & 0.50 & ~1$\times$10$^{-3}$
                    & & & \cite{Selvadurai2015,Xia2022,Rodrigues2020,Azad2005,Paul_Blessington_Selvadurai_2015}
                    &
                    0.5 & n
                    & \cite{Tariq2022}
                    \\
                    \cline{4-5}
                    \cline{8-10}
                    & Mg & 0.30 & \multicolumn{2}{c|}{\multirow{2}{*}{/}}
                    &
                    & \cite{Triyono2020}
                    & \multicolumn{3}{c|}{}
                    \\
                    & Co & 0.10 & \multicolumn{2}{c|}{}
                    &
                    & \cite{Gu2021}
                    & \multicolumn{3}{c|}{}
                    \\
                    \cline{3-10}
                    & Nb
                    & \multicolumn{5}{c|}{\multirow{1}{*}{}}
                    & 0.25
                    & n
                    & \cite{Zhou2021}
                    \\
                    \cline{9-9}
                    & V
                    & \multicolumn{5}{c|}{\multirow{1}{*}{}}
                    & 0.25
                    & y
                    & \cite{Azouzi2021}
                    \\
                    \cline{1-10}
                    \multirow{7}{*}{Bi(Fe$_{1-x}$D$_x$)O$_3$}
                    & Co & 0.30 & ~5$\times$10$^{-2}$
                    & \multirow{4}{*}{RT}
                    & \multirow{5}{*}{y}
                    & \cite{Khajonrit2018,Sui2015,Kharel2008} & 0.125
                    & \multirow{3}{*}{y}
                    & \cite{Lu2017}
                    \\
                    & Mn & 0.20 & /
                    & &
                    & \cite{Sosnowska2002} & 0.125
                    &
                    & \cite{Lu2017}
                    \\
                    & Cr, Ni, V & 0.03 & ~1$\times$10$^{-3}$
                    & &
                    & \cite{Kharel2008} & 0.125
                    &
                    & \cite{Lu2017}
                    \\
                    \cline{8-10}
                    & Nb & 0.01 & ~1$\times$10$^{-3}$
                    & &
                    & \cite{Jun2005}
                    & \multicolumn{3}{c|}{}
                    \\
                    \cline{4-5}
                    & Y & 0.10 & \multicolumn{2}{c|}{/} &
                    & \cite{Mukherjee2012}
                    & \multicolumn{3}{c|}{}
                    \\
                    \cline{3-10}
                    & Cu, Zn
                    & \multicolumn{5}{c|}{}
                    & 0.25 &
                    y &
                    \cite{Rong2016}
                    \\
                    \cline{1-10}
                    \multirow{5}{*}{(Ni$_{1-x}$D$_x$)O}
                    & Zn & 0.05 & 1$\times$10$^{-4}$
                    & \multirow{4}{*}{RT}
                    & \multirow{4}{*}{y}
                    & \cite{Panigrahi2020}
                    & \multicolumn{3}{c|}{}
                    \\
                    & Fe & 0.02 & 1$\times$10$^{-4}$
                    & &
                    & \cite{Manna2008}
                    & \multicolumn{3}{c|}{}
                    \\
                    & Mn & 0.06 & 1$\times$10$^{-3}$
                    & &
                    & \cite{Layek2016}
                    & \multicolumn{3}{c|}{}
                    \\
                    & Nd & 0.03 & 1$\times$10$^{-4}$
                    & &
                    & \cite{Rahman2018}
                    & \multicolumn{3}{c|}{}
                    \\
                    \cline{3-10}
                    & Li, Cu, Ag
                    & \multicolumn{5}{c|}{}
                    &
                    0.083 &
                    y & \cite{Egbo2020}
                    \\
                    \cline{1-10}
                    \multirow{2}{*}{(Mn$_{1-x}$D$_x$)Te}
                    & Cu & 0.075 & /
                    & RT
                    & \multirow{3}{*}{/}
                    & \cite{Ren2016}
                    & \multicolumn{3}{c|}{}
                    \\
                    \cline{4-5}
                    & Cr & 0.05 & ~3$\times$10$^{-2}$
                    & 280 K
                    &
                    & \cite{Polash2020}
                    & \multicolumn{3}{c|}{}
                    \\
                    \cline{1-5}
                    Sr(Tc$_{1-x}$D$_x$)O$_3$ & Ru & 0.75
                    & ~4$\times$10$^{-2}$
                    & 150 K
                    &
                    & \cite{Reynolds_2020}
                    & \multicolumn{3}{c|}{}
                    \\
                    \hline
                \end{tabular}
            }}
        \label{dop_exp}
    \end{table*}

    \section{spin-dependent doping}

    \begin{figure}[hbpt]
        \centering
        \includegraphics[scale=0.3,angle=0]{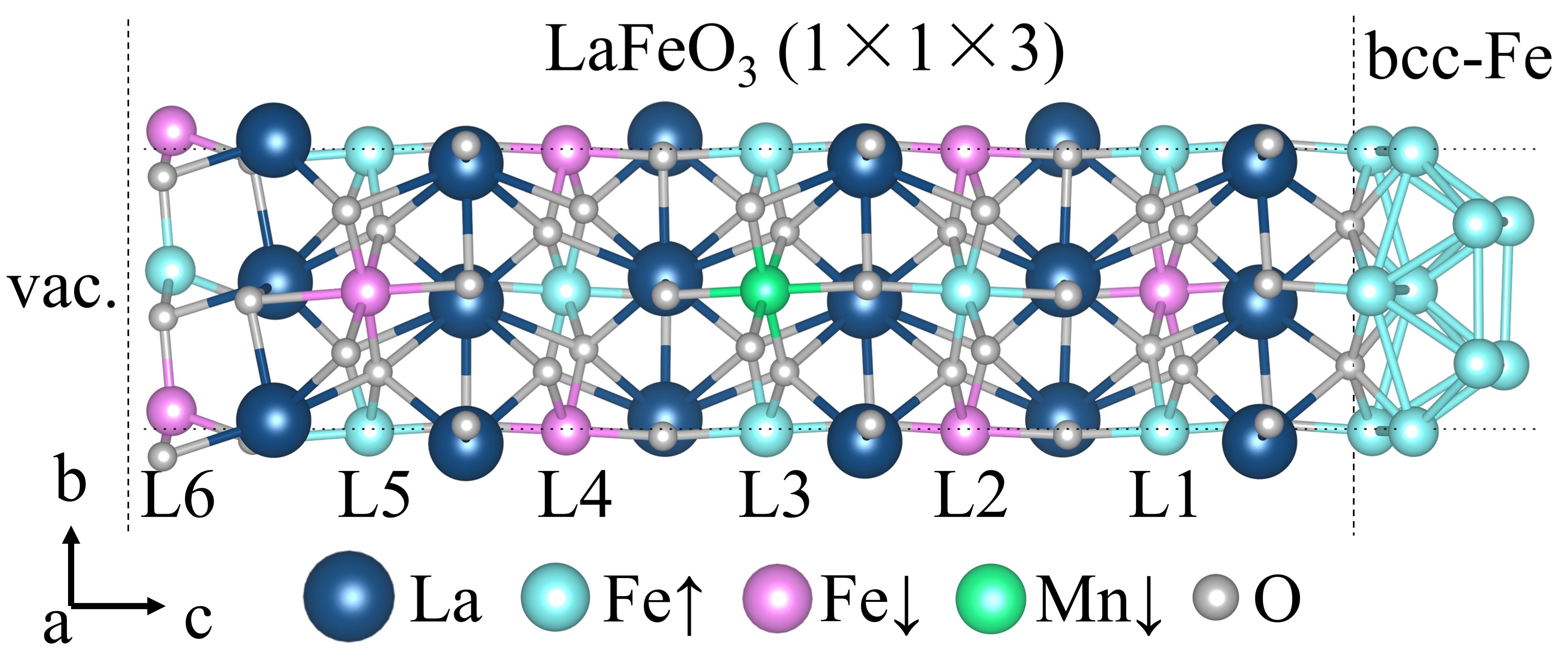}\\
        \caption{Crystal structure of the supercell La(Fe, Mn)O$_3$/bcc-Fe, where the Mn impurity are doped at Fe site of layer L3.}\label{fig6}
    \end{figure}
    LaFeO$_3$ has a G-AFM ground state, and shows very weak ferromagnetism due to the spin canting caused by the Dzyaloshinskii-Moriya (DM) interaction \cite{Dzyaloshinsky1958}.
    The net magnetic moment per Fe atom in LaFeO$_3$ is about 10$^{-4} \mu_B$.
    Experiments found that doping at Fe sites will increase the net magnetic moment to 10$^{-4} \sim 10^{-2}$ $\mu_B$ per Fe atom, while it's still in the G-AFM state, as shown in Tab. \ref{dop_exp}.

    To break the symmetry of spin up and down sublattices in LaFeO$_3$, we study the AFM insulator LaFeO$_3$ with FM substrate of Fe metal, and consider a LaFeO$_3$/bcc-Fe heterojunction, as shown in Fig.~\ref{fig6}.
    The lattice constant is a = 2.87 \AA~for bcc-Fe, and a = 5.60 \AA, b = 5.66 \AA~for LaFeO$_3$.
    The lattice of 2$\times$2$\times$1 bcc-Fe and LaFeO$_3$ fit well with a small lattice mismatch about 1\%.
    The optimized lattice constants of LaFeO$_3$/bcc-Fe heterojunction are a = b = 5.56 \AA, where three layers of LaFeO$_3$, one layer of bcc-Fe along (001) direction, and a vacuum layer of 20 \AA ~are considered.

    For simplicity, we fix the spin of bcc-Fe substrate as spin up.
    The total energy difference of the supercells La(Fe,Mn)O$_3$/bcc-Fe with Mn at spin up and down sublattices is shown in Table \ref{table3}.
    For the doped Mn at Fe positions of L2, L3, L4 layers, the impurities Mn tend to occupy the positions of spin down sublattice.
    This spin-dependent doping process makes La(Fe,Mn)O$_3$ into FIM state with large net magnetic moment.

    \begin{table}[btht]
        \setlength{\tabcolsep}{2mm}
        \caption{{
            Total energy difference of the supercells La(Fe,Mn)O$_3$/bcc-Fe with Mn at spin up and down sublattices.
            Layers L2 to L4 are defined in Fig. \ref{fig6}}.}
        {
            \scalebox{1}

            {
                \begin{tabular}{c|c|c|c}
                    \hline
                    \hline
                    \multirow{2}{*}{\makebox[0.10\textwidth][c]{Position of Mn}} &
                    \multirow{1}{*}{\makebox[0.09\textwidth][c]{Distance to}} &
                    \multirow{2}{*}{\makebox[0.09\textwidth][c]{Ground state}} &
                    \multirow{1}{*}{\makebox[0.09\textwidth][c]{E$_\uparrow$ - E$_\downarrow$}}
                    \\
                    &
                    \multirow{1}{*}{interface (nm)} &
                    &
                    \multirow{1}{*}{(meV)}
                    \\
                    \hline
                    L2 & 0.8 & \multirow{3}{*}{FIM} & 64.1 \\
                    L3 & 1.2 &                      & 53.1 \\
                    L4 & 1.6 &                      & 88.6 \\
                    \hline
                    \hline

                \end{tabular}
            }}
        \label{table3}
    \end{table}

    The spin-dependent doping can be explained by the effective magnetic field provided by the substrate of FM bcc-Fe.
    The effective magnetic field from Fe substrate reduces the energy of Layer $\alpha$ by $\sum_{i}^{i\in \alpha}\left(\vec{S_i}\cdot\vec{H}^\alpha_{eff} \right)$, where $i$ represents the magnetic atoms in layer $\alpha$,  $\vec{H}^\alpha_{eff}$ is the effective magentic field at layer $\alpha$ from the Fe substrate, $\vec{S_i}$ is the magnetic moment of atom $i$.
    Because $\vec{H}^\alpha_{eff}$ from Fe substrate is in direction of spin up, there is a competition between Mn and Fe for spin up sublattice, and Fe always win because of its bigger magnetic moment (4.15 $\mu_B$) compared with Mn (3.73 $\mu_B$), resulting in the impurities Mn occupying spin down position.
    The energy difference of the supercells La(Fe,Mn)O$_3$/bcc-Fe with Mn at spin up and down sublattices is still significant when impurities Mn are doped at the layer 4, i.e., 1.6 nm to the interface.
    Since the LaFeO$_3$ nanosheets could be as thin as 5 nm \cite{Gao2020,Yu2022}, the influence of Fe substrate is effective.
    The spin-dependent doping will lead to spin polarization of dopants and induce AFM-FIM transition.
    Experiment found that the magnetic field will significantly increase the net magnetic moment of ZnO doped with 2\% Cr \cite{Li2010a}.

    \section{Magnetic properties of FIM Semiconductors}

    \subsection{$T_C$ in LaFeO$_3$-based FIM semiconductors}
\begin{figure}[!bht]
        \centering
        \includegraphics[scale=0.28,angle=0]{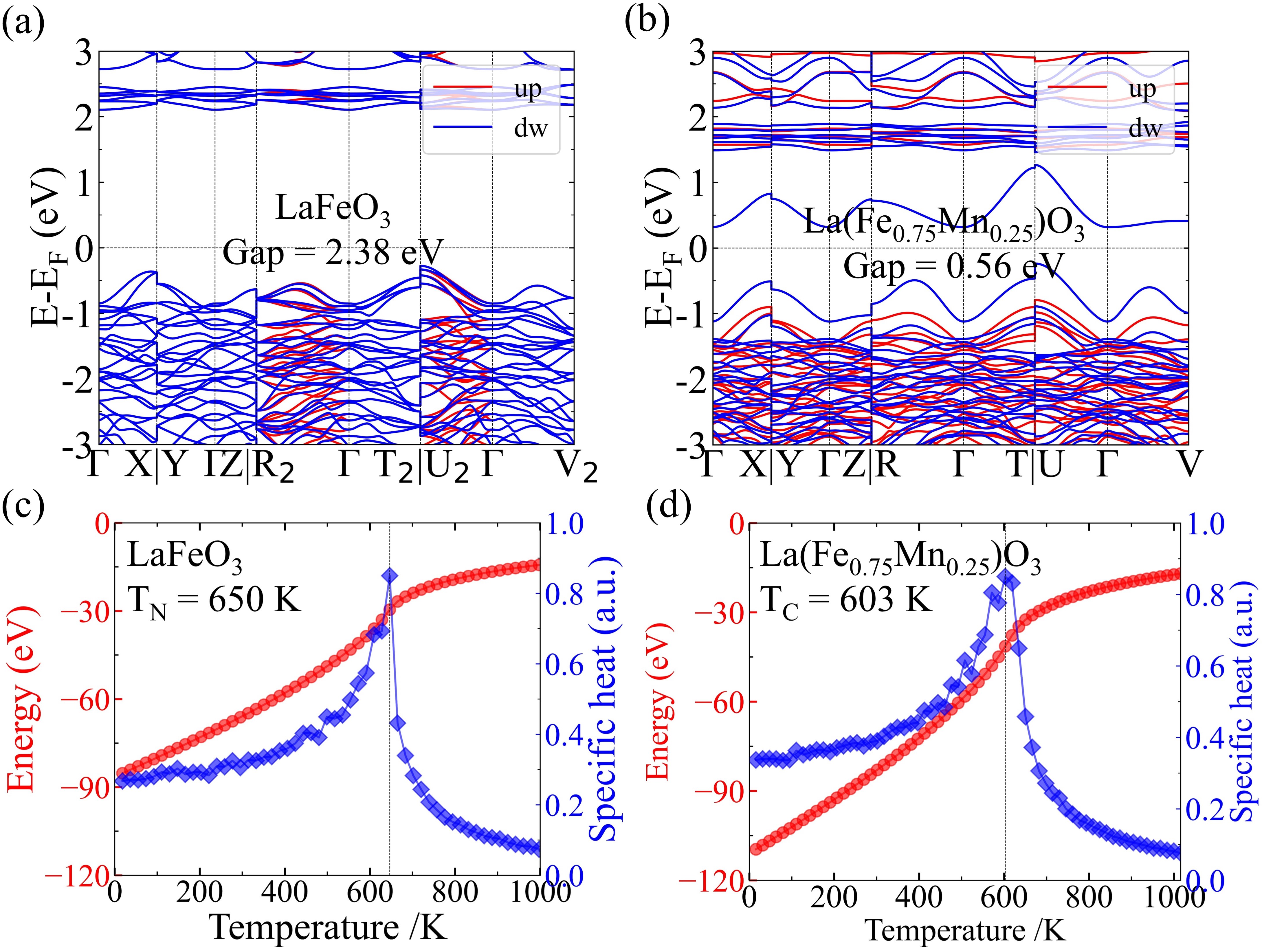}\\
        \caption{
            DFT results of band structure for (a) LaFeO$_3$ with a band gap of 2.38 eV and (b) La(Fe$_{0.75}$Mn$_{0.25}$)O$_3$ with a band gap of 0.56 eV.
            Monte Carlo results of energy and specific heat as a function of temperature for (c)
            LaFeO$_3$ with Neel temperature $T_N$ = 650 K and (d) La(Fe$_{0.75}$Mn$_{0.25}$)O$_3$ with Curie temperature $T_C$ = 603 K.
        }\label{fig2}
    \end{figure}

    The band structure of LaFeO$_3$ is shown in Fig \ref{fig2} (a), with a band gap of 2.38 eV, consistent with the experimental value of 2.05$\sim$2.51 eV \cite{Sasikala2017,Manzoor2018}.
    Since LaFeO$_3$ is AFM with zero net magnetic moment, we determine its $T_N$ through energy and specific heat by Monte Carlo simulations.
    The results are shown in Fig. \ref{fig2} (c), with a sharp peak of specific heat at $T_N$ = 650 K, close to the experimental value of 740 K \cite{Koehler1957}.

    For the La(Fe$_{0.75}$Mn$_{0.25}$)O$_3$ where one of the four Fe atoms is replaced by a Mn atom in a LaFeO$_3$ unitcell.
    DFT results show that its magnetic ground state is FIM.
    Mn has a magnetic moment of 3.73 $\mu_B$, smaller than Fe (4.18 $\mu_B$), induce a net magnetic moment near 0.12 $\mu_B$ per LaFeO$_3$ unitcell.
    In addition, La(Fe$_{0.75}$Mn$_{0.25}$)O$_3$ is a FIM semiconductor with a band gap of 0.56 eV, and a high Curie temperature $T_C$ = 603 K is estimated by the Monte Carlo simulation, as shown in Fig. \ref{fig2} (d).

\begin{figure}[thb]
        \centering
        \includegraphics[scale=0.27,angle=0]{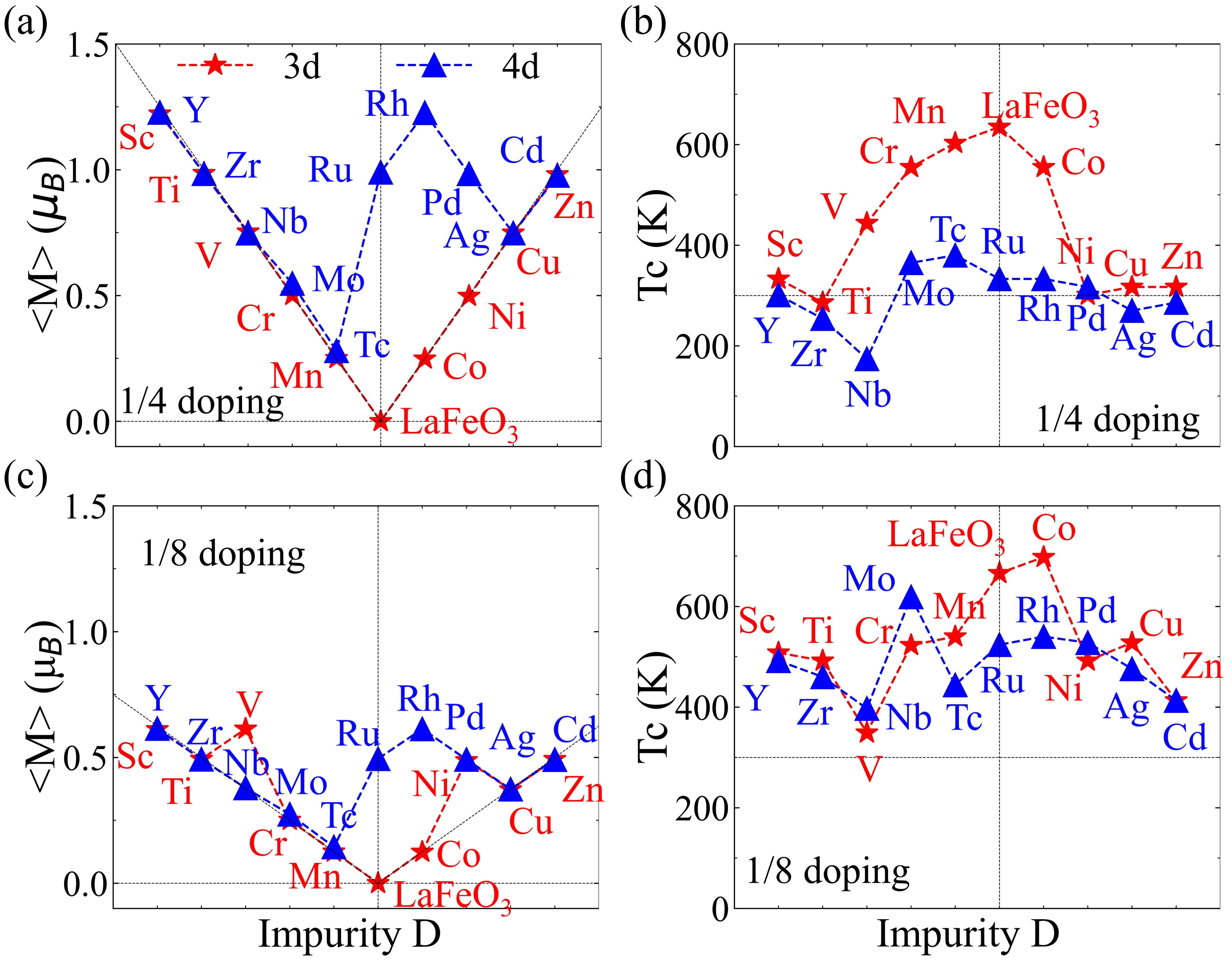}\\
        \caption{
            (a) Average magnetic moment per magnetic atom $<$M$>$ and (b) Curie temperature $T_C$ for La(Fe$_{0.75}$D$_{0.25}$)O$_3$.
            (c) $<$M$>$ and (d) Tc for La(Fe$_{0.875}$D$_{0.125}$)O$_3$.
            The impurity D is taken as 3d and 4d transition metal elements. For comparison, the $T_N$ = 650 K of host LaFeO$_3$ is also included in (b) and (d).
        }\label{fig3}
    \end{figure}

    With different 3d and 4d dopants, the magnetic ground states of La(Fe$_{0.75}$D$_{0.25}$)O$_3$ maintain FIM.
    Because the magnetic moments of Fe are almost constant compared with different dopants, the net magnetic moment are from the broken of the symmetry of the AFM spin sublattices, which can be calculated as M$_{tot}$= $|$M$_{dopant}$ - M$_{Fe}|$, the detailed magnetic moments see Supplemental Material \cite{SM}.
    The average magnetic moment per lattice $\langle M \rangle$ of La(Fe$_{0.75}$D$_{0.25}$)O$_3$ is defined as $\langle M \rangle =M_{tot}/N$, the magnetic lattice number N = 4 for the LaFeO$_3$ unitcell, and the results are shown in Fig. \ref{fig3} (a).
    The Curie temperature $T_C$ of La(Fe$_{0.75}$D$_{0.25}$)O$_3$ which was estimated by the Monte Carlo simulations, as shown in Fig. \ref{fig3}(b).
    It is noted that most of $T_C$ with 3d and 4d dopants are above room temperature.

    To discuss the effect of concentrations, the material La(Fe$_{0.875}$D$_{0.125}$)O$_3$ is studied.
    A 2$\times$1$\times$1 supercell is considered, where one of eight Fe atoms is replaced by the D (3d or 4d) atom.
    DFT results show that its magnetic ground state maintain FIM with different dopants.
    The $<$M$>$ of La(Fe$_{0.875}$D$_{0.125}$)O$_3$ is about half to that of La(Fe$_{0.75}$D$_{0.25}$)O$_3$, as shown in Fig. \ref{fig3}(c). It is expected since the concentration of dopants decreases from 1/4 to 1/8.
    It is interesting to note that the $T_C$ of La(Fe$_{0.875}$D$_{0.125}$)O$_3$ are higher than that of La(Fe$_{0.75}$D$_{0.25}$)O$_3$, as shown in Fig. \ref{fig3} (d).
    The calculated values of average magnetic moment per lattice $<$M$>$, Curie temperature Tc, and band gap of La(Fe$_{0.75}$D$_{0.25}$)O$_3$ and La(Fe$_{0.875}$D$_{0.125}$)O$_3$ are summarized in Tab. \ref{table2}.

    \begin{table*}[thb]
        \setlength{\tabcolsep}{1.5mm}
        \caption{The calculated results of the average magnetic moment per magnetic atom $<$M$>$, band gap, and Curie temperature $T_C$ for La(Fe$_{0.75}$D$_{0.25}$)O$_3$ and La(Fe$_{0.875}$D$_{0.125}$)O$_3$.
        }
        {
            \scalebox{1}

            {
                \begin{tabular}{cc|ccc|ccc|l}
                    \hline
                    \hline
                    \multicolumn{2}{c|}{\multirow{2}{*}{
                    		\makebox[0.20\textwidth][c]{\diagbox{Dopants}{Properties}}
                    	}} &
                    \multicolumn{3}{c|}{\makebox[0.21\textwidth][c]{La(Fe$_{0.75}$D$_{0.25}$)O$_3$}} &
                    \multicolumn{3}{c|}{\makebox[0.21\textwidth][c]{La(Fe$_{0.875}$D$_{0.125}$)O$_3$}}&
                    \multirow{2}{*}{\makebox[0.10\textwidth][c]{Experiments}}
                    \\
                    \cline{3-8}
                    & &
                    \makebox[0.07\textwidth][c]{$<$M$>$ ($\mu_B$)} &
                    \makebox[0.07\textwidth][c]{Gap (eV)}
                     &
                    \makebox[0.07\textwidth][c]{$T_C$ (K)}
                     &
                    \makebox[0.07\textwidth][c]{$<$M$>$ ($\mu_B$)}
                     &
                    \makebox[0.07\textwidth][c]{Gap (eV)}
                     &
                    \makebox[0.07\textwidth][c]{$T_C$ (K)}
                     &
                    \\
                    \hline
                    & Sc & 1.22 & 2.23 & 333 & 0.61 & 2.32 & 508 &                                                                                            \\
                    & Ti & 0.99 & 1.32 & 286 & 0.49 & 1.17 & 492 & Ref. \cite{Phokha2015,Sasikala2018,Sasikala2017}                                           \\
                    & V  & 0.75 & 1.51 & 444 & 0.61 & 0.12 & 349 &                                                                                            \\
                    & Cr & 0.50 & 2.06 & 555 & 0.25 & 2.28 & 523 & Ref. \cite{Selvadurai2015,Xia2022,Rodrigues2020,Azad2005,Paul_Blessington_Selvadurai_2015} \\
                    3d     & Mn & 0.25 & 0.56 & 603 & 0.13 & 0.98 & 540 &                                                                                            \\
                    atoms  & Co & 0.25 & 1.48 & 555 & 0.12 & 1.45 & 698 & Ref. \cite{Gu2021}                                                                         \\
                    doping & Ni & 0.50 & 0.64 & 301 & 0.49 & 0.40 & 492 & Ref. \cite{Idrees2011}                                                                     \\
                    & Cu & 0.75 & 0.54 & 317 & 0.37 & 0.65 & 528 & Ref. \cite{Dogdibegovic2016}                                                               \\
                    & Zn & 0.98 & 0.26 & 317 & 0.49 & 0.00 & 413 & Ref. \cite{Manzoor2018,Bhat2013}                                                           \\
                    \hline\hline
                    & Y  & 1.23 & 1.98 & 301 & 0.61 & 2.26 & 492 &                                                                                            \\
                    & Zr & 0.98 & 1.23 & 254 & 0.49 & 1.27 & 460 &                                                                                            \\
                    & Nb & 0.75 & 1.65 & 174 & 0.38 & 0.00 & 397 &                                                                                            \\
                    & Mo & 0.55 & 0.62 & 365 & 0.27 & 0.70 & 619 & Ref. \cite{Jana2019}                                                                       \\
                    4d     & Tc & 0.28 & 1.13 & 380 & 0.14 & 0.00 & 444 &                                                                                            \\
                    atoms  & Ru & 0.99 & 0.96 & 333 & 0.49 & 1.07 & 524 &                                                                                            \\
                    doping & Rh & 1.23 & 1.30 & 333 & 0.61 & 1.60 & 540 &                                                                                            \\
                    & Pd & 0.98 & 0.00 & 317 & 0.49 & 0.56 & 528 &                                                                                            \\
                    & Ag & 0.75 & 0.34 & 270 & 0.37 & 0.57 & 476 &                                                                                            \\
                    & Cd & 0.98 & 0.45 & 286 & 0.49 & 0.00 & 413 &                                                                                            \\
                    \hline
                    \hline

                \end{tabular}
            }}

        \label{table2}
    \end{table*}

    \subsection{MOKE in LaFeO$_3$-based FIM semiconductors}
    \begin{figure}[hbpt]
        \centering
        \includegraphics[scale=0.4,angle=0]{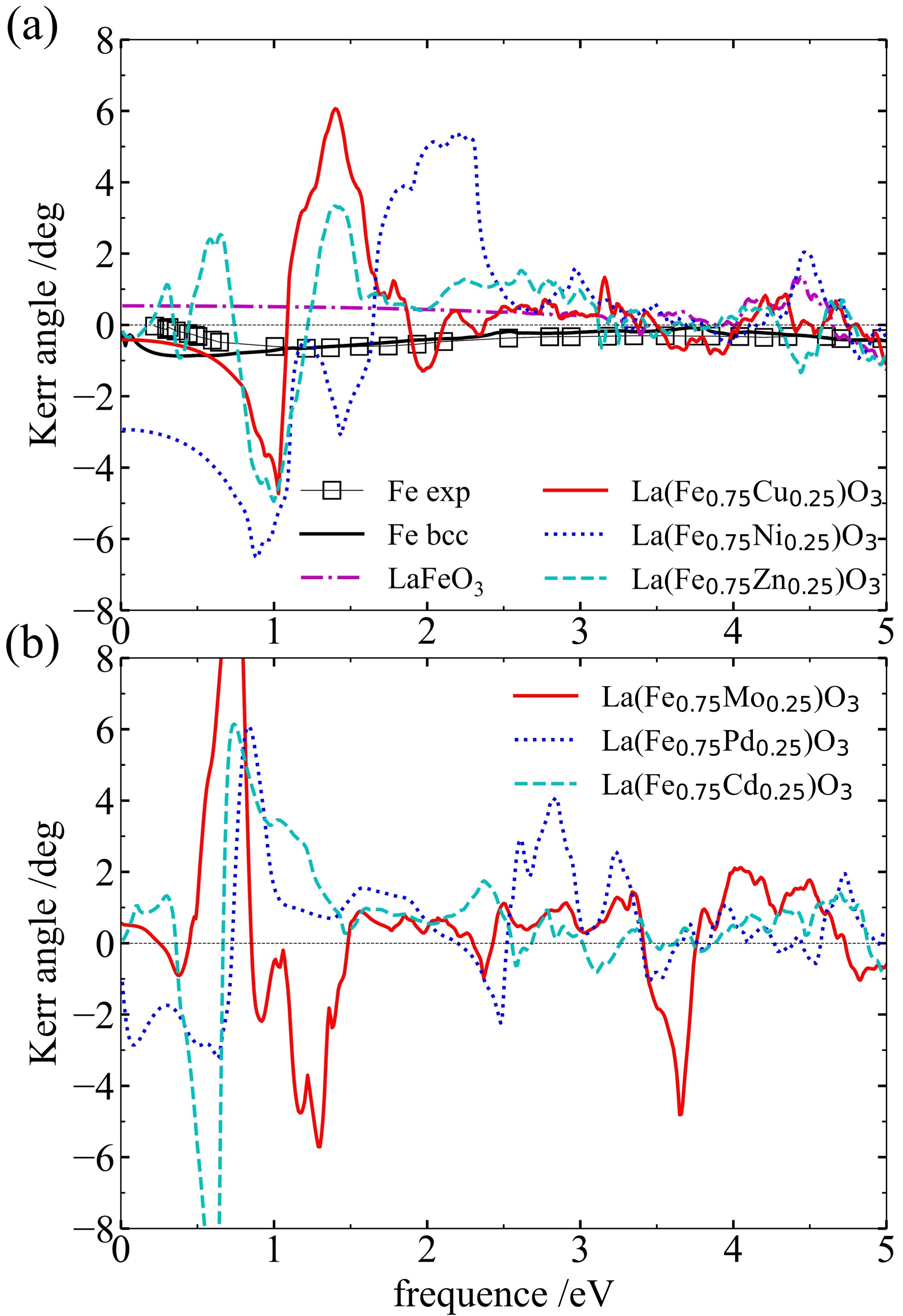}\\
        \caption{
            (a) DFT results of Kerr angle for Fe, LaFeO$_3$, and La(Fe$_{0.75}$D$_{0.25}$)O$_3$ with D = Ni, Cu, and Zn.
            (b) DFT results of Kerr rotation angle for La(Fe$_{0.75}$D$_{0.25}$)O$_3$ with D = Mo, Pd, and Cd.
            Experimental Kerr rotatoin angle of Fe \cite{Oppeneer1992} is also included for comparison.  }\label{fig4}
    \end{figure}
    We investigated the magneto-optical Kerr effect for La(Fe$_{0.75}$D$_{0.25}$)O$_3$.
    The Kerr rotation angle is given by:
    \begin{eqnarray}
        \begin{aligned}
            \theta_K(\omega)=Re\frac{\varepsilon_{xy}}{(1-\varepsilon_{xx})\sqrt{\varepsilon_{xx}}},
            \\
        \end{aligned}
        \label{eqs1}
    \end{eqnarray}
    where where $\varepsilon_{xx}$ and $\varepsilon_{xy}$ are the diagonal and off-diagonal components of the dielectric tensor $\varepsilon$, $\omega$ is the frequency of incident light.
    The dielectric tensor $\varepsilon$ can be obtained by the optical conductivity tensor $\sigma$ as $\varepsilon(\omega)=\frac{4{\pi}i}{\omega}\sigma(\omega)+I$, where I is the unit tensor.
    The calculated $\varepsilon(\omega)$ as a function of photon energy for LaFeO$_3$, and La(Fe$_{0.75}$D$_{0.25}$)O$_3$ with D = Ni, Cu, Zn, Mo and Pd is shown in Fig. \ref{fig4}.
    The experimental result for Fe \cite{Oppeneer1992} and our DFT result for Fe bulk are also included for comparison.
    There are a big Kerr angle for La(Fe$_{0.75}$D$_{0.25}$)O$_3$ with $\omega$ $<$ 2 eV, about 10 times bigger than bcc Fe.
    It is worth noting that LaFeO$_3$ shows small but non-zero Kerr angle, despite its collinear AFM order, this may be related to the room temperature ferroelectricity of LaFeO$_3$ \cite{Acharya2010}.
    Detailed results of Kerr angle are given in Supplemental Material \cite{SM}.

    \begin{table*}[hpt]

        \setlength{\tabcolsep}{1.8mm}
        \caption{
            The calculated band gap and $T_N$ for some high $T_N$ AFM insulators and semiconductors with chemical formula ABO$_3$, and the calculated band gap, $T_C$ and $<$M$>$ for their doped materials A(B$_{0.75}$D$_{0.25}$)O$_3$.
            The impurity D is taken as some 3d and 4d transition metal elements.
        }
        \begin{tabular}{ccc|cccc|c}
            \hline
            \multicolumn{3}{c|}{\makebox[0.21\textwidth][c]{Host ABO$_3$}} & \multicolumn{4}{c|}{
            	\makebox[0.28\textwidth][c]{A(B$_{0.75}$D$_{0.25}$)O$_3$}
            	}
            	&
            	\makebox[0.10\textwidth][c]{Experiments} 
            \\ 
            \hline
            \makebox[0.07\textwidth][c]{Material}
            & \makebox[0.07\textwidth][c]{Gap (eV)}
            & \makebox[0.07\textwidth][c]{$T_N$ (K)}
            & \makebox[0.07\textwidth][c]{D}
            & \makebox[0.07\textwidth][c]{Gap (eV)}
            & \makebox[0.07\textwidth][c]{$T_C$ (K)}
            & \makebox[0.07\textwidth][c]{$<$M$>$ ($\mu_B$)}
            & \makebox[0.10\textwidth][c]{Ref}
            \\ \hline
            \multirow{5}{*}{LaFeO$_3$} & \multirow{5}{*}{2.4} & \multirow{5}{*}{650}
            & V & 1.51 & 444 & 0.75
            &
            \\ & & & Cr & 2.06 & 555 & 0.50
            & \cite{Selvadurai2015,Xia2022,Rodrigues2020,Azad2005,Paul_Blessington_Selvadurai_2015}
            \\ & & & Co & 1.48 & 555 & 0.25& \cite{Gu2021}
            \\ & & & Mo & 0.62 & 365 & 0.55& \cite{Jana2019}
            \\ & & & Ru & 0.96 & 333 & 0.99 &
            \\ \hline
            \multirow{5}{*}{BiFeO$_3$} & \multirow{5}{*}{2.3} & \multirow{5}{*}{580}
            & V & 1.61 & 397 & 0.75 & \cite{Kharel2008}
            \\ & & & Cr & 1.96 & 524 & 0.50 & \cite{Kharel2008}
            \\ & & & Co & 2.00 & 476 & 0.25 & \cite{Khajonrit2018,Sui2015,Kharel2008}
            \\ & & & Mo & 0.69 & 333 & 0.55 &
            \\ & & & Ru & 0.94 & 602 & 0.99 &
            \\ \hline
            \multirow{5}{*}{SrTcO$_3$} & \multirow{5}{*}{1.5} & \multirow{5}{*}{883} & V & 0.84 & 793 & 0.47 &
            \\ & & & Cr & 0.00 & 634 & 0.25 &
            \\ & & & Co & 0.13 & 476 & 0.51 &
            \\ & & & Mo & 0.12 & 555 & 0.25 &
            \\ & & & Ru & 0.50 & 635 & 0.24 & \cite{Reynolds_2020}
            \\ \hline
            \multirow{5}{*}{CaTcO$_3$} & \multirow{5}{*}{1.5} & \multirow{5}{*}{587}
            & V & 0.95 & 482 & 0.47 &
            \\ & & & Cr & 0.00 & 355 & 0.25 &
            \\ & & & Co & 0.14 & 343 & 0.51 &
            \\ & & & Mo & 0.09 & 393 & 0.25 &
            \\ & & & Ru & 0.56 & 444 & 0.24 &
            \\ \hline
        \end{tabular}
        \label{table4}
    \end{table*}

    \subsection{Other high $T_C$ FIM semiconductors}
    In addtion to LaFeO$_3$, we also study the doping of other high $T_N$ AFM insulators and semiconductors, including BiFeO$_3$, SrTcO$_3$, CaTcO$_3$.
    The calculation results are shown in Table \ref{table4}.
    When 25\% of the 3d transition metal element of host are replaced by other 3d or 4d impurities, many room temperature FIM semiconductors are obtained in LaFeO$_3$, BiFeO$_3$, SrTcO$_3$ and CaTcO$_3$.
    All of these host materials are perovskite with $T_N$ above 550 K and band gap bigger than 1.5 eV.
    Detailed results are given in Supplemental Material \cite{SM}.
    For the same impurity and concentration, $T_C$ and band gap obtained after doping are positively related to $T_N$ and band gap of AFM material.
    According to the calculation results, room temperature FIM semiconductors could be obtained by doping in AFM semiconductors, and a high $T_N$ and a large band gap are needed.

    \subsection{Mean-field theory of the effect of doping on $T_C$}
    To study the influence of different impurities on $T_C$, as shown in Fig. \ref{fig3}, we use the Weiss molecular field approximate \cite{Dai2017x}.
    By the simple AFM Heisenberg model and the mean-field approximation (MFA), we get $T_N$ of G-AFM LaFeO$_3$ as
    \begin{equation}
        \begin{aligned}
            T_N=2\frac{J_0 S_0(S_0+1)}{k_B},
        \end{aligned}
        \label{eq1}
    \end{equation}
    where $J_0$ represents the nearest-neighbor coupling constant of Fe-Fe in LaFeO$_3$, $S_0$ is the magnetic moment of Fe in LaFeO$_3$, and $k_B$ is the Boltzmann constant.
    By the help of DFT calculation, $J_0$ = 2.25 meV, $S_0$ = 4.15 $\mu_B$.
    By Eq.\eqref{eq1}, it has $T_N$ = 1115 K.
    It is noted that the $T_N$ = 1115 K by mean-field theory of Eq.\eqref{eq1} is much higher than the $T_N$ = 650 K by the Monte Carlo simulation with the same $J_0$ and the $T_N$ = 740 K of LaFeO$_3$ in experiment \cite{Koehler1957}.

    By the similar mean-field theory, we can obtain the expression of $T_C$ for FIM semiconductors La(Fe,D)O$_3$.
    For simplicity, we only discuss the case of one impurity per unitcell without disorder, and only the nearest-neighbor coupling are considered.

    The ratio of $T_C$ and $T_N$ is expressed as:

    \begin{equation}
        \begin{aligned}
            \frac{T_C}{T_N}
            &=t_0\sqrt{\frac{a+\sqrt {a^2-b}}{8}},
            \\
            a&=\frac{1}{9}
            \left[
            6(6-z_{AB})t_D+z_{AB}z_{BA}+6(6-z_{BA})
            \right],
            \\
            b&=\frac{16}{9}t_D(6-z_{AB})(6-z_{BA}),\\
            t_0&=\frac{J_1}{J_0}\frac{S(S+1)}{S_0(S_0+1)},
            t_D=\left( \frac{J_2}{J_1} \right)^2 \frac{S_D(S_D+1)}{S(S+1)},
        \end{aligned}
        \label{eq2}
    \end{equation}
    where $J_0$, $J_1$ are the nearest-neighbor coupling constants of Fe-Fe in LaFeO$_3$ and La(Fe,D)O$_3$, respectively, $J_2$ is the nearest-neighbor coupling constants between Fe and D in La(Fe,D)O$_3$.
    $S_0$, $S$ are the magnetic moments of Fe in LaFeO$_3$ and La(Fe,D)O$_3$, respectively, and $S_D$ is the magnetic moment of D in La(Fe,D)O$_3$, $z_{ij}$ is the coordination number of the site j near the site i.
    Supposing dopants at spin down sites, sublattice A mean Fe atoms spin up with nearest-neighbor impurities, sublattice B mean Fe atoms spin down without nearest-neighbor impurities, respectively.
    Here $t_0$ describes the ratio of Fe-Fe couplings in La(Fe,D)O$_3$ and LaFeO$_3$, $t_D$ describes the ratio of Fe-D coupling and Fe-Fe coupling in La(Fe,D)O$_3$.
    See detailed information in Supplemental Material \cite{SM}.

    \begin{figure}[hbpt]
        \centering
        \includegraphics[scale=0.29,angle=0]{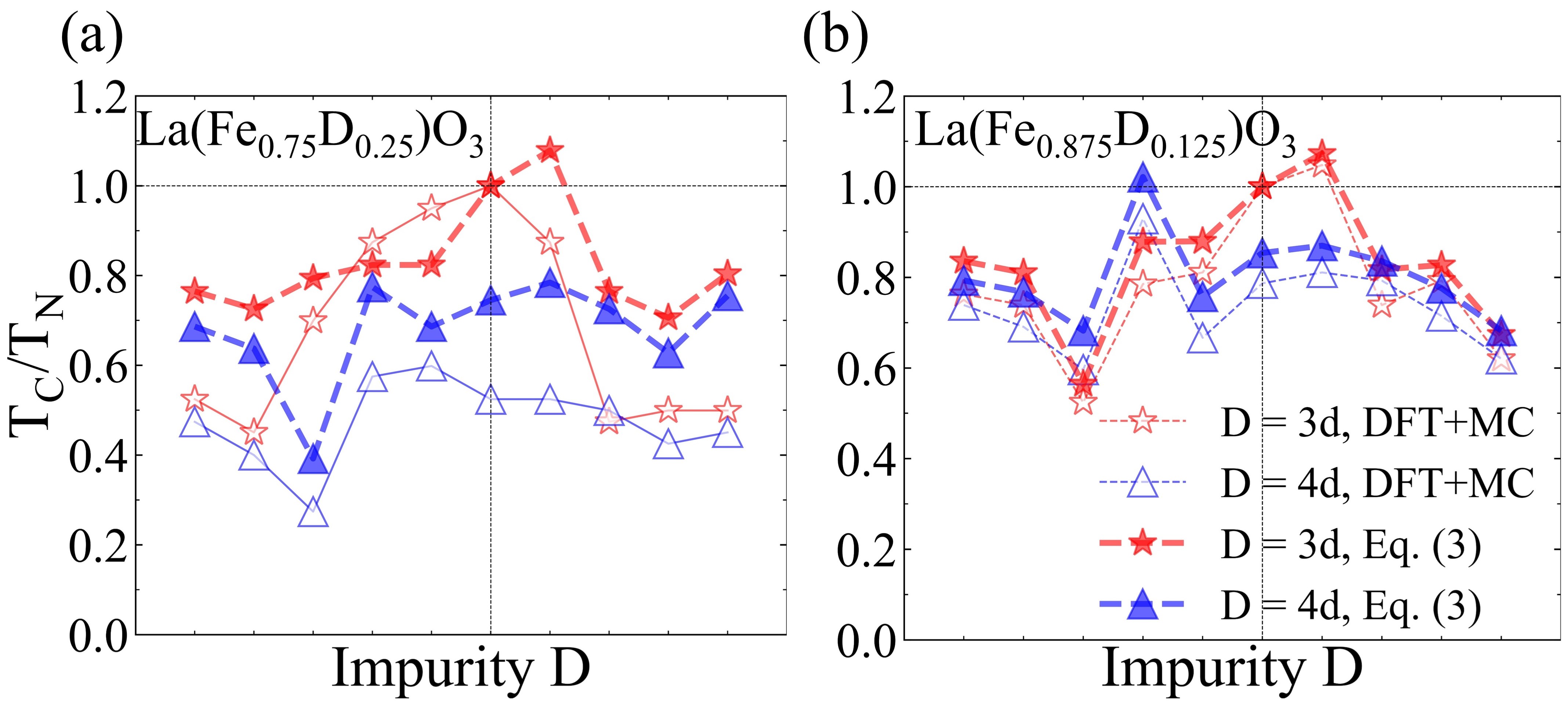}\\
        \caption{
            For $T_N$ of LaFeO$_3$ and $T_C$ of La(Fe$_{1-x}$D$_{x}$)O$_3$, the ratio of $T_C$/$T_N$ for (a) x = 0.25 and (b) x = 0.125.
            The impurity D is taken as 3d and 4d transition metal elements.
            The numerical results (DFT+MC) are taken from Figs. \ref{fig3} (b) and (d).
            The mean-field approximation results are obtained by Eq. \eqref{eq2}.
        }\label{fig5}
    \end{figure}

    For case of 1/4 doping, the coordination number is $z_{AB}$ = 4, $z_{BA}$ = 6.
    For case of 1/8 doping, the coordination number is $z_{AB}$ = 4, $z_{BA}$ = 4.
    Take these parameters and coupling constant and magnetic moment from DFT into Eq. \eqref{eq2}, we obtain the ratio of $T_C/T_N$ for La(Fe$_{0.75}$D$_{0.25}$)O$_3$ and La(Fe$_{0.875}$D$_{0.125}$)O$_3$, as shown in Figs. \ref{fig5} (a) and (b), respectively.
    The ratio of $T_C/T_N$ obtained by Eq. \eqref{eq2} with the mean-field approximation (MFA) and numerical calculations (DFT+MC) shown in Fig. \ref{fig3} are in a good agreement.
    Thus, we note that it is possible to understand the effect of doping on $T_C$ in FIM semiconductors La(Fe,D)O$_3$ by the Eq. \eqref{eq2} of the conventional mean-field theory.

    \section{conclusion}
    Based on the DFT calculations, we show an approach to obtain room temperature FIM semiconductors by spin-dependent doping in high $T_N$ insulators and semiconductors with large band gap.
    To demonstrate the spin-dependent doping, the Mn-doped AFM insulator LaFeO$_3$ with FM sublattices bcc-Fe is studied by the DFT calculation.
    It is shown that the doped Mn impurities prefer to occupy one sublattice of LaFeO$_3$ due to the effective magnetic field of substrate bcc-Fe, and obtain the FIM semiconductor La(Fe,Mn)O$_3$ with large magnetic moment.
    By this method, we predict a series of room temperature FIM semiconductors in La(Fe,D)O$_3$, where D denoted dopant of 3d and 4d transition metals.
    Large magneto-optical Kerr effect were found in La(Fe$_{0.75}$D$_{0.25}$)O$_3$.
    By the equation of mean-field approximation, the ration of $T_C$ in La(Fe,D)O$_3$ and $T_N$ of LaFeO$_3$ are obtained, in a good agreement with the numerical results of DFT + MC.
    In the same way, the FIM semiconductors with high $T_C$ are also predicted in some other high $T_N$ AFM insulators and semiconductors, such as BiFeO$_3$, SrTcO$_3$, CaTcO$_3$, etc.
    Our results suggest that the spin-dependent doping is a promising way to produce high $T_C$ FIM semiconductors from high $T_N$ AFM insulators and semiconductors.

    \section* {Acknowledgements}
    This work is supported by the National Key R\&D Program of China (Grant No. 2022YFA1405100), National Natural Science Foundation of China (Grant No. 12074378), the Beijing Natural Science Foundation (Grant No. Z190011), National Natural Science Foundation of China (Grant No. 11834014), the Beijing Municipal Science and Technology Commission (Grant No. Z191100007219013), the Chinese Academy of Sciences Project for Young Scientists in Basic Research (Grant No. YSBR-030), and the Strategic Priority Research Program of Chinese Academy of Sciences (Grants No. XDB28000000 and No. XDB33000000).

    \bibliographystyle{apsrev4-2}

\end{document}